\documentclass[aps,pra,twocolumn,amsmath,amssymb,superscriptaddress]{revtex4-1}

\usepackage{appendix}
\usepackage{ mathrsfs }
\usepackage{hyperref}
\usepackage{graphicx}

\usepackage{dsfont}
\usepackage{physics}
\usepackage{wrapfig}
\usepackage{epstopdf}
\usepackage{pifont}
\usepackage{amsmath}
\usepackage{amssymb}

\usepackage{verbatim}

\begin{document}


\title{Entanglement assisted state discrimination and entanglement preservation}


\author{\"{O}zen\c{c} G\"{u}ng\"{o}r}

\email[]{oxg34@case.edu}
\affiliation{Department of Physics, Case Western Reserve University,
10900 Euclid Avenue, Cleveland, Ohio 44106, USA}
\affiliation{Department of Physics, Middle East Technical University, 06800, Ankara, Turkey}

\author{Sadi Turgut}
\affiliation{Department of Physics, Middle East Technical University, 06800, Ankara, Turkey}


\date{\today}

\begin{abstract}
In this paper, the following scenario is considered: there are two qubits
possessed by two parties at different locations. Qubits have been prepared
in one of a maximum of four, mutually-orthogonal, entangled states and the
parties wish to distinguish between the states by using local operations
and classical communication. Although in general it is not possible to
distinguish between four arbitrary states, the parties can spend some
pre-shared entanglement to achieve perfect discrimination between four
qubit states and can also preserve the entanglement of the states after
discrimination. This is shown by employing the theory of majorization and
the connections between entanglement transformations and state
discrimination protocols.
\end{abstract}


\maketitle

\section{Introduction}
One of the consequences of quantum mechanics is that it is impossible to
perfectly distinguish between non-orthogonal pure quantum states. Much work
has been done in this field and several procedures have been identified for
minimizing the probability of error when discriminating between pure,
monopartite quantum states \cite{chefles, Ivan, Peres, Dieks}. While not as
general, partial results and necessary conditions on the separability of
monopartite, mixed quantum states exist in the literature \cite{generalproof,
chefcond}.

While any number of monopartite, orthogonal states can be perfectly distinguished, this is not true in general for multipartite, entangled states under local operations and classical communication (LOCC). Walgate\cite{walgate} has shown that two, orthogonal, bipartite states can be perfectly distinguished using LOCC alone and Virmani and his collaborators\cite{Virmani} have extended this result into non orthogonal states. Bandyopadhyay and his collaborators have shown that the number of states that are perfectly distinguishable under LOCC is closely related to the entanglement content of the states \cite{bandyodisc, bandyodisc2} and they have found a general bound on the number
\begin{equation}
N\leq \frac{D}{\overline{1+R(\ket{\psi_{i}})}} \leq \frac{D}{\overline{2^{E_{R}(\ket{\psi_{i}})}}} \leq \frac{D}{\overline{2^{E_{g}(\ket{\psi_{i}})}}}
\end{equation}
where $R(\ket{\psi_{i}})$ \cite{robustness} is the global robustness of entanglement, $E_{R}(\ket{\psi_{i}})$ is the relative entanglement entropy \cite{relativeent}, $E_{g}(\ket{\psi_{i}})$ is the geometric measure of entanglement \cite{geometricentanglement} for the state $\ket{\psi_{i}}$ and $D$ is the dimension. An explicit example of 4 Bell type states being indistinguishable can be found in the work of Ghosh et. al. \cite{4state} .

The results listed above by various authors have one aspect in common: the
entanglement in the states is lost after discrimination, the parties are left
with a product state in their hands. Cohen has considered the possibility of
preserving entanglement after discrimination in his paper \cite{cohenent} by
considering the Schmidt ranks of the states to be discriminated as a measure
of the entanglement without explicitly referring to the Schmidt coefficients
themselves. He has also shown that entanglement can be used as a resource in
state discrimination procedures \cite{cohenresource} enabling the parties to
achieve tasks that are otherwise impossible.

In this paper, our main point is to argue that the following set of mutually
orthogonal Bell type states
\begin{equation}
\label{states}
\begin{split}
\ket{\psi_{1}} &= a\ket{00}+b\ket{11} \\
\ket{\psi_{2}} &= b^{*}\ket{00}-a^{*}\ket{11} \\
\ket{\psi_{3}} &= c\ket{01}+d\ket{10} \\
\ket{\psi_{4}} &= d^{*}\ket{01}-c^{*}\ket{10}
\end{split}
\end{equation}
are distinguishable under LOCC if the parties agree to spend some pre-shared
entanglement and the parties can achieve entanglement preservation after
discrimination by also spending pre-shared entanglement. To prove these
claims, we first view multipartite state discrimination as a local
entanglement transformation and use the theory of majorization
\cite{majlong,maj1} to classify these transformations. These ideas also serve
to show the usefulness of majorization relations for characterizing local
state discrimination procedures.

Majorization defines a partial order between vectors whose components sum up
to the same value. It is defined as follows. Let $x$ and $y$ be $d$
dimensional vectors
\begin{equation}
x=\begin{pmatrix}
x_{1} \\
x_{2} \\
\vdots \\
x_{d}
\end{pmatrix}, \, \, y=\begin{pmatrix}
y_{1} \\
y_{2} \\
\vdots \\
y_{d}
\end{pmatrix}.
\end{equation}
Then, $y$ majorizes $x$, or $x \prec y$, if the following inequality holds
for every $k = 1,2,\dots,d$ and equality holds for $k=d$
\begin{equation}
\sum_{i=1}^{k}x_{i}^{\downarrow} \leq \sum_{i=1}^{k}y_{i}^{\downarrow},
\end{equation}
where $x^{\downarrow}$ denotes the vector obtained by organizing the
components of $x$ in descending order. Specifically, $x_{i}^{\downarrow}$ is
the $i^{th}$ largest component of $x$. If the dimensions of the two vectors
are not equal, the lower dimensional one can be ``padded'' with enough zeros.
Majorization relations can also be extended to matrices of equal trace by
constructing a vector $\lambda(A)$ where the elements are the eigenvalues of
$A$ \emph{sorted in a decreasing order}. Nielsen has shown that for two
entangled states in Schmidt forms
\begin{equation}
\begin{split}
\ket{\psi}&=\sum_{i}\sqrt{x_{i}}\ket{ii}\\
\ket{\phi}&=\sum_{i}\sqrt{y_{i}}\ket{i^{\prime}i^{\prime}}
\end{split}
\end{equation}
where $x_{i}$ and $y_{i}$ are the respective Schmidt coefficients, which both
sum up to 1, the state $\ket{\psi}$ can be converted into $\ket{\phi}$ under
LOCC with unit probability if and only if $\lambda(\psi)\prec \lambda(\phi)$
(or $x\prec y$) is satisfied where $\lambda(\psi)=x$ is defined to be the
eigenvalue vector of the reduced density matrix
$\rho_{A}=\tr_{B}\dyad{\psi}$. Subsequently, Jonathan and Plenio
\cite{JonathanPlenio}
have shown that the state $\ket{\psi}$ can be converted into a set of
different states where $\ket{\phi_{i}}$ is obtained with probability $p_{i}$,
if and only if
\begin{equation}
\lambda(\psi) \prec \sum_{i}p_{i}\lambda(\phi_{i}).
\end{equation}
The above equation is of central importance to our discussion on the distinguishability of entangled quantum states as it establishes the connection between entangled state transformations and state discrimination.

The majorization relations described above don't explicitly describe how to
construct the local operators to realize the transformation but the
construction of the majorization relation guarantees the existence of a LOCC
measurement scheme. In the case of a deterministic transformation, the
operation is equivalent to one of the parties making a general measurement
and the other party using a unitary transformation conditional on the outcome
of the general measurement. In the case of  transformations for mixed bipartite or pure multipartite states, a one-way transformation is not enough.

The above relation can be used in the following way to investigate the
conditions for the distinguishability of the states in Eq.~\eqref{states}.
Suppose Alice, Bob, Charlie and Devin share the pure state
\begin{equation}\label{state4}
\ket{\Psi}=\frac{1}{2}\sum_{i=1}^{4}\ket{\psi_{i}}_{AB}\ket{\phi_{i}}_{CD}
\end{equation}
where $\ket{\phi_{i}}$ are the Bell states and we have chosen that the
initial probabilities $p_{i}$ of the states $\ket{\psi_{i}}$ to be equal. If
Alice and Bob are able to distinguish between the states $\ket{\psi_{i}}$,
Charlie and Devin will share the Bell state $\ket{\phi_{i}}$ with probability
$p_{i} = 1/4$. Distinguishing between the states $\ket{\psi_{i}}$ in this
case is equal to transforming the state $\ket{\Psi}$ into $\ket{\phi_{i}}$
with $p_{i}=1/4$ with $i$ identified meaning that Alice and Bob have
succeeded in discriminating the states $\ket{\psi_{i}}$. If $\ket{\Psi}$ is
reinterpreted as a bipartite state where Alice and Charlie together is
considered as one of the parties (and Bob and Devin is the other party), the
following majorization relation can be constructed for this entanglement
transformation, after padding the eigenvalue matrix of $\frac{1}{2} \mathds{1}_{2}$ with enough zeros
\begin{equation}
  \label{maj1}
  \lambda(\Psi) \prec \begin{pmatrix}
  1/2 \\ 1/2 \\ 0 \\ 0
 
  \end{pmatrix},
\end{equation}
leading to the inequality
\begin{equation}
  \label{ineq1}
  \frac{1}{8}(a+b+c+d)^{2} \leq \frac{1}{2} ~,
\end{equation}
where we have assumed, without losing any generality, that $a,b,c,d$ are real
and $a\geq b$ and $c\geq d$. Since
$x +\sqrt{1-x^{2}} \geq 1$ for $1\geq x \geq 0$,
the inequality in Eq.~\eqref{ineq1} can only be satisfied if both $a,c=1$
which is the case that the states $\ket{\psi_{i}}$ are product states. This
serves to show that the 4 orthogonal states $\ket{\psi_{i}}$ cannot be
perfectly discriminated if they are entangled.

The same procedure can be applied to investigate the distinguishability of 3
of the four states in Eq.~(\ref{states}) in the following way. Consider the
state shared between Alice, Bob, Charlie and Devin
\begin{equation}
\ket{\Psi}=\frac{1}{\sqrt{3}}\sum_{i=1}^3\ket{\psi_{i}}_{AB}\ket{\phi_{i}}_{CD}.
\end{equation}
Using the majorization relation in Eq.~(\ref{maj1}), we find that if two of
the three states are maximally entangled, discrimination is not possible but
numerical calculations show that discrimination is possible for a range of
values for the Schmidt parameters $a$ and $c$. The conditions for
discrimination found by majorization relations are in complete agreement with
the results in Ref.~\onlinecite{4state}.

\section{Entanglement Assisted Discriminaton}
Majorization relations are also helpful in entanglement assisted
discrimination. The situation that is considered is as follows. Alice and Bob
have qubits in an unknown state among $\ket{\psi_i}$, but they also have
other qubits in a known entangled state $\ket{\phi}$. Can they successfully
achieve discrimination with an incurred cost of consuming the entanglement in
$\ket{\phi}$? In other words, can they discriminate between the states
$\ket{\phi}\otimes\ket{\psi_i}$? The entangled state to be used up will be
denoted by
\begin{equation}\label{bank}
\ket{\phi}=\alpha \ket{00}+\beta \ket{11}
\end{equation}
where the important parameters are the Schmidt parameters $\alpha$ and $\beta$.

\begin{figure}[h]
\begin{center}

\includegraphics[width=8cm]{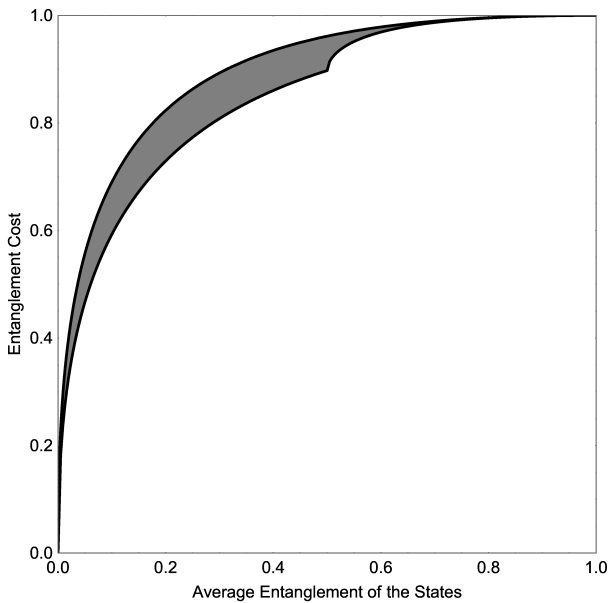}
\caption{Plot of the entanglement cost of discrimination versus the average entanglement of the states $\ket{\psi_{i}}$ for $p_{i}=1/4$. The entanglement cost is not a simple function of the average entanglement as the average entanglement is not a one-to-one function of the parameters $a$ and $c$.\label{fig:avent4nl}}
\end{center}

\end{figure}

The same idea used before to characterize entangled state discrimination will
be applied here; if Alice and Bob are able to distinguish between the states
$\ket{\phi}\otimes\ket{\psi_i}$, then when they apply the same protocol to
$\ket{\phi} \otimes \ket{\Psi}$, the parties Charlie and Devin will be left
with a Bell state $\ket{\phi_{i}}$, with probability $p_{i}=1/4$ for each.
This is equivalent to a probabilistic entanglement transformation of the
state $\ket{\phi}\otimes \ket{\Psi}$ to $\ket{\phi_{i}}$ with probability
distribution $p_{i}$. Using this idea, the following majorization relation
can be constructed
\begin{equation}
 \label{maj2}
 \lambda(\phi) \otimes \lambda(\Psi) \prec \lambda(\frac{1}{2}\mathds{1})~.
\end{equation}
The following bound for the Schmidt coefficients of the state $\ket{\phi}$
can be obtained from this relation
\begin{equation}\label{ineq2}
\abs{\alpha}^{2} \leq \frac{4}{(a+b+c+d)^2}~,
\end{equation}
where it is assumed that $\abs{\alpha}^{2} \geq \abs{\beta}^{2}$. The
entanglement entropy for the state $\ket{\phi}$ can be calculated for various
entanglement of the states $\ket{\psi_{i}}$ using the regular expression for
entanglement entropy
\begin{equation}
\label{ententropy}
\mathcal{E}(\phi)=-\sum_{i=1}^{2}\lambda_{i} \log_{2}\lambda_{i}
\end{equation}
where $\lambda_{i}$ are the Schmidt coefficients of the state $\ket{\phi}$
and  the entanglement needed to discriminate the states $\ket{\psi_{i}}$
versus their average entanglement is shown in Fig.~\ref{fig:avent4nl}.

The parties can also succeed in discriminating between the states
$\ket{\psi_{i}}$ in the following way: one of the parties can teleport
her 
part of the states to the other party using an e-bit of entanglement and the
states can be locally discriminated in a perfect way afterwards. The cost
calculated using majorization, however, shows that there are protocols for
which the entanglement cost can be lower than the teleportation cost provided
that all the states are not maximally entangled.

\section{Discrimination with Remaining Entanglement}

In this section we turn our attention to discrimination of entangled states with
preservation of their entanglement. We use entanglement entropy as before to
quantify entanglement and use the fact that a bipartite pure state of qubits
is convertible to another one with the same or smaller entanglement entropy
under LOCC. This enables us to characterize the discrimination protocol by considering only the Schmidt coefficients of the states since two states with
the same Schmidt coefficients are same up to local unitaries.

We consider two observers, Alice and Bob, in possession of qubits AB whose
state is one of the states $\ket{\psi_{i}}$ as in Eq.~(\ref{states}).
They also possess quantum systems A$'$B$'$
with a known but arbitrary entangled state $\ket{\phi}$ whose entanglement is
usable by the parties. An absolute upper bound on the entanglement cost of
this protocol can be found by considering the cost of using quantum
teleportation. Alice can teleport her part of the states to Bob using up one
e-bit of entanglement after which Bob can perfectly discriminate between 4
orthogonal quantum states. Bob then teleports to Alice what was originally
her part of the state using up another e-bit of entanglement bringing up the
total cost to 2 e-bits of entanglement. In other words, with 2 e-bits of
entanglement of A$'$B$'$, the task can always be accomplished.

To obtain a better bound on the amount of entanglement necessary to
accomplish the task, the discrimination procedure will be used to design an
entanglement transformation protocol. Suppose that the parties are able to
discriminate between the states $\ket{\psi_{i}}$ of qubits AB and preserve
the states, with the aid of the other pair A$'$B$'$ in state $\ket{\phi}$. In
other words, there is a local procedure implemented by the parties such that
if the pair AB is initially in state $\ket{\psi_i}_{AB}$, then the procedure
causes the transformation
\begin{equation}
  \label{eq:TransformationWithPreserving}
  \ket{\phi}_{A'B'}\otimes\ket{\psi_i}_{AB} \longrightarrow \ket{00}_{A'B'}\otimes\ket{\psi_i}_{AB}
\end{equation}
and the parties identify the index $i$. (In here and the following, the
symbols A, A$'$, $\overline{\mathrm{A}}$, etc. represent particles possessed
by Alice and, similarly, B, B$'$, $\overline{\mathrm{B}}$ belong to Bob.)
Now, we consider a situation where four particles
AB$\overline{\mathrm{A}}\overline{\mathrm{B}}$ are in state
\begin{equation}
  \label{Psi}
  \ket{\Psi}_{AB\overline{A}\overline{B}}=\sum_i
      \ket{\psi_{i}}_{AB}\otimes \ket{\varphi_{i}}_{\overline{A}\,\overline{B}}
\end{equation}
and the same procedure is applied. This then causes a probabilistic
transformation
\begin{equation}
  \label{transformationreal}
  \ket{\phi}_{A'B'}\otimes \ket{\Psi}_{AB\overline{A}\overline{B}} \longrightarrow
       \ket{00}_{A'B'}
       \otimes\ket{\psi_i}_{AB}
       \otimes\ket{\varphi_{i}}_{\overline{A}\overline{B}}
\end{equation}
where the $i$th outcome is identified and obtained with probability
$p_i=\Vert\varphi_i\Vert^2$. The majorization relation will then put on
bounds on the possibility of the original transformation in
\eqref{eq:TransformationWithPreserving}. Note that there are infinitely many
such bounds because $\ket{\varphi_i}$ can be chosen arbitrarily. However, the
best bounds are obtained when $\ket{\Psi}$ is a product state in the
A$\overline{\mathrm{A}}$:B$\overline{\mathrm{B}}$ cut because
\begin{equation}
\label{productstatement}
\lambda(\Psi)=\begin{pmatrix}
1 \\ 0
\end{pmatrix}~.
\end{equation}
If $\ket{\Psi}$ is an entangled state, largest entry in the Schmidt
coefficients will be smaller than 1; this will give us looser bounds for
$\ket{\phi}$.

\begin{figure}[h]
\begin{center}
\includegraphics[width=8cm]{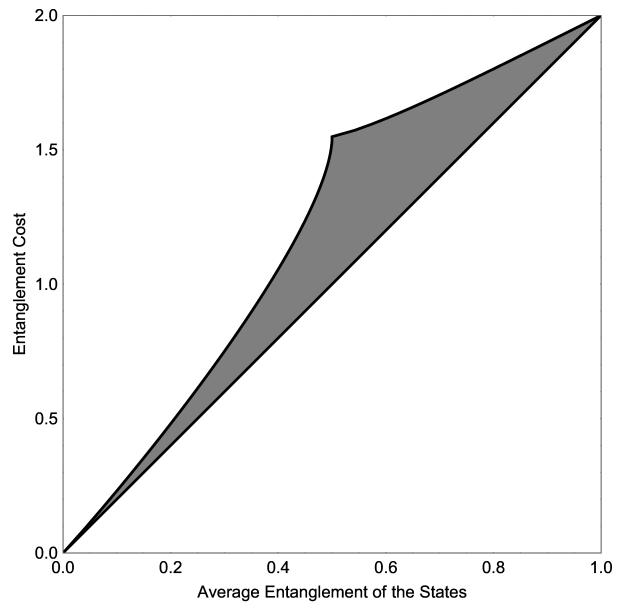}
\caption{Plot of the entanglement cost versus the average entanglement
  of the states. The diagonal line is when all the states have equal
  entanglement and the cusp is when two of the states are maximally
  entangled and the others are product states.}
\label{fig:avent4cost}
\end{center}

\end{figure}

When we choose $\ket{\Psi}$ to be a product state, the majorizarion relation
that characterizes the transformation and therefore gives us bounds on
$\ket{\phi}$ can be expressed as
\begin{equation}
  \label{majorizationent}
  \lambda(\phi) \prec \sum_{i}\norm{\varphi_{i}}^{2}\lambda(\psi_{i}) \otimes \lambda\left(\frac{\varphi_{i}}{\norm{\varphi_{i}}}\right).
\end{equation}
The state $\ket{\Psi}$, which is unentangled in the
A$\overline{\mathrm{A}}$:B$\overline{\mathrm{B}}$ cut, can be expressed as
$\ket{\Psi}=\ket{u}_{A\overline{A}}\otimes \ket{v}_{B\overline{B}}$.
We choose $\ket{u}$ and $\ket{v}$ as Bell states
\begin{equation}
  \label{u,v}
  \ket{u} = \ket{v}= \frac{1}{\sqrt{2}}(\ket{00}+\ket{11})
\end{equation}
and re-express $\ket{\Psi}$ as
\begin{equation}
\label{Psi2}
\ket{\Psi}=\frac{1}{2}\sum_{i,j,k,l=0}^{1}\delta_{ij}\delta_{kl}\ket{ik}_{AB}\otimes \ket{jl}_{\bar{A}\bar{B}}.
\end{equation}
We define, for the ease of manipulation, what we call a partial inner product
between bras and kets in the following way
\begin{equation}
  \label{partialproduct}
  \ket{\varphi_{m}}=\braket{\psi_{m}}{\Psi}\rangle
\end{equation}
and express $\ket{\psi_{m}}$ as
$\ket{\psi_{m}}=\sum_{i,k}(\psi_{m})_{ik}\ket{ik}_{AB}$. The states
$\ket{\varphi_{m}}$ can then be found as
\begin{equation}
  \label{varphi}
  \begin{split}
  \ket{\varphi_{m}}=\braket{\psi_{m}}{\Psi}\rangle &= \frac{1}{2}\sum_{i,j,k,l}(\psi_{m})^{\star}_{ik}\delta_{ij}\delta_{kl}\ket{jl}_{\bar{A}\bar{B}} \\  \ket{\varphi_{m}} &= \frac{1}{2}\ket{\psi^{\star}_{m}}.
  \end{split}
\end{equation}
where $\ket{\psi^{\star}_{m}}$ is the state obtained by complex conjugating
the coefficients of $\ket{\psi_{m}}$ in the ``computational basis''. The
majorization relation can now be expressed as
\begin{equation}
\label{majfin}
\begin{split}
\lambda(\phi) &\prec \frac{1}{4} \sum_{i}\lambda(\psi_{i}) \otimes \lambda(\psi_{i}), \\
\lambda(\phi) &\prec \frac{1}{2}
\begin{pmatrix} a^{4} + c^{4}
\\ a^{2}b^{2}+c^{2}d^{2}
\\ a^{2}b^{2}+c^{2}d^{2}
\\ b^{4} + d^{4}
\end{pmatrix}
\end{split}
\end{equation}
where we assumed $a\geq b$, $c\geq d$ and $a,b,c,d$ are real without losing
any generality. The majorization relation in Eq.~(\ref{majfin}) shows that
the state $\ket{\phi}$ has at least 4 Schmidt coefficients, and it has 2
e-bits of entanglement if all the $\ket{\psi_{i}}$ are maximally entangled,
and no entanglement if all the $\ket{\psi_{i}}$ are product states. This
result is in agreement and is expected since the upper bound on the
entanglement cost of such a protocol is 2 e-bits if teleportation is used. The
entanglement cost versus the average entanglement of the states is plotten in
Fig.~\ref{fig:avent4cost}

\section{Conclusion}
In this work, we have considered the problem of local discrimination of
multipartite states and shown that majorization relations prove very useful
in characterizing various types of state discrimination problems. Using
majorization relation, it is possible to say something about the minimum
amount of entanglement that must be used to discriminate between 4
orthogonal, bipartite, entangled states. Bounds have been provided on the
entanglement that needs to be spent. These results are in agreement with the
teleportation bound and the cost that we have calculated is always lower than
or equal to the teleportation bound.

The question of preserving entanglement after discrimination is also
considered. This problem has also been discussed by Cohen \cite{cohenent},
and the protocols have been examined. Using majorization relations for
probabilistic entanglement transformations, we have found bounds on the
Schmidt coefficients of the resource state and calculated the entanglement
cost of entanglement preserving discrimination problems. The upper bound on
the cost is in agreement with the teleportation cost of 2 e-bits and is always
smaller than 2 e-bits for non maximally entangled states.

We haven't constructed explicit protocols that would achieve discrimination
for the given problems but as it is stated above, the majorization relations,
by construction, guarantee the existence of a LOCC protocol. There is however, a caveat, the protocols constructed from the majorization relations are local protocols in the sense that Alice can act on the Hilbert spaces $A, \overline{A}$. In the case of a protocol achieving discrimination with remaining entanglement, Alice can't act on the Hilbert space $\overline{A}$ at all. Thus, an entanglement transformation protocol on 4 particles $A,\overline{A},B,\overline{B}$ can be constructed for the discrimination problem of 2 particles $A,B$ but not the other way around. In the light of this identification, the bounds calculated in this work are necessary conditions for parties to achieve discrimination with remaining entanglement. Sufficient conditions for the existence of such protocols and their explicit construction is still an open question.

\bibliography{arxiv_ver1.bib}
\bibliographystyle{unsrt}


\end{document}